<Content Type: Article>

[Title] **Segmented readout for Cherenkov time-of-flight positron emission tomography detectors based on bismuth germanate**

[Name] Minseok Yi, Daehee Lee, Alberto Gola, Stefano Merzi, Michele Penna, Jae Sung Lee, Simon R. Cherry, Sun Il Kwon

**Abstract**

Positron emission tomography (PET) is the most sensitive biomedical imaging modality for non-invasively detecting and visualizing positron-emitting radiopharmaceuticals within a subject. In PET, measuring the time-of-flight (TOF) information for each pair of 511-keV annihilation photons improves effective sensitivity but requires high timing resolution. Hybrid materials that emit both scintillation and Cherenkov photons, such as bismuth germanate (BGO), recently offer the potential for more precise timing information from Cherenkov photons while maintaining adequate energy resolution from scintillation photons. However, a significant challenge in using such hybrid materials for TOF PET applications lies in the event-dependent timing spread caused by the mixed detection of Cherenkov and scintillation photons due to relatively lower production of Cherenkov photons. This study introduces an innovative approach by segmenting silicon photomultiplier (SiPM) pixels coupled to a single crystal, rather than using traditional SiPMs that are as large as or larger than the crystals they read. We demonstrated that multiple time stamps and photon counts obtained from the segmented SiPM can classify events by providing temporal photon density, effectively addressing this challenge. The approach and findings would lead to new opportunities in applications that require precise timing and photon counting, spanning the fields of medical imaging, high-energy physics, and optical physics.

**Main**

Positron emission tomography (PET) has revolutionized medical imaging by providing non-invasive, quantitative measurements of physiological processes *in vivo*. The fundamental principle of PET relies on detecting paired annihilation photons emitted indirectly by a positron-emitting radionuclide, which is introduced into the body as part of a biologically active molecule. Recent advancements in PET instrumentation, particularly the development of whole-body and total-body time-of-flight (TOF) PET systems, represent a major leap forward in medical imaging technology. These systems offer unprecedented opportunities for comprehensive disease assessment, pharmacokinetic studies, and early detection of metastases **[1-4]**. The extended axial field of view (AFOV) of up to approximately 200 cm allows for simultaneous imaging of multiple organs, opening new avenues for studying complex diseases and systemic conditions **[5, 6]**.

While these advanced TOF-PET systems provide remarkable benefits, they also require substantial resources in terms of detector materials, electronics, and data processing capabilities. To address these challenges and further enhance PET performance, bismuth germanate (BGO) crystals have recently garnered renewed interest due to their potential as hybrid scintillator/Cherenkov emitters. The use of BGO crystals in such systems could potentially make whole-body and total-body PET more economically viable for widespread clinical adoption while still benefiting from the excellent sensitivity to the high energy annihilation photons and hybrid emission properties of BGO **[7-12]**. The dual-emission characteristic of BGO offers the possibility of improved timing resolution through the exploitation of the prompt Cherenkov signal, while maintaining the high detection efficiency associated with its scintillation properties **[13, 14]**. This unique combination of performance and affordability positions BGO as a compelling candidate for next-generation PET instrumentation, potentially enabling faster TOF capabilities, improved image quality, and more accessible advanced PET imaging technologies.

In addition to their rapid generation within a few picoseconds (ps) after an annihilation photon interaction, the Cherenkov photons exhibit several distinct characteristics that set them apart from scintillation photons. For instance, the number of Cherenkov photons generated from a 511 keV annihilation photon interaction is relatively low, totaling less than 30, in contrast to approximately 4,000 scintillation photons in BGO **[15, 16]**. Moreover, Cherenkov photons are predominantly produced in the UV/blue spectrum, whereas scintillation photons are emitted across a wavelength range from 400 to 700 nm **[13, 17]**. Therefore, numerous studies have been conducted with the aim of advancing silicon photomultipliers (SiPMs) and high-frequency electronics, with a specific emphasis on enhancing the detection efficiency of Cherenkov photons while preserving their prompt nature **[13, 18-27]**.

One of the major challenges in making use of a mixture of Cherenkov and scintillation photons in BGO is the event-dependent timing spread introduced by the low Cherenkov photon statistics and its high fluctuations in its detection count. Therefore, if it is possible to classify whether an event was triggered by Cherenkov photons or scintillation photons and additionally determine the temporal density of promptly detected photons, it would allow for the use of an appropriate TOF kernel for each event in the image reconstruction process **[28-30]**. This, in turn, would significantly enhance the image signal-to-noise ratio **[31]**.

In this study, we hypothesized that if two independent readouts from a single annihilation photon event produce two individual timestamps, along with the number of initially detected photons at each readout, the proximity of these timestamps would indicate a higher initial photon density for that event. The key innovation in this work is to segment the SiPM pixel being coupled to a single crystal which is the reverse of the traditional approach where SiPM pixels are equal to or larger in area than the crystals they read out. In doing so, we can also leverage the secondary benefits inherent to the characteristics of the SiPM detector itself. In the context of Cherenkov radiation detection for fast timing, it is crucial to accurately and precisely detect the prompt Cherenkov photons due to their ultra-fast emission characteristics **[16]**. Achieving the necessary precision in this detection is challenging, as it requires a level of accuracy approaching individual photon detections. Thus, as opposed to scintillation light, which is characterized by photon-rich conditions, enhancing the timing performance requires greater emphasis on achieving larger and faster single photon response. When addressing this challenge, it becomes evident that the higher number of parasitic capacitances connected in parallel in SiPM detectors may cause a low-pass filtering effect that could degrade timing performance. Recently, in collaboration with Fondazione Bruno Kessler (FBK), a new type of SiPM named OctaSiPM was developed. The OctaSiPM is a SiPM array consisting of pixels with small active areas, reducing the intrinsic parasitic capacitance of the detector. In this study, we couple a 3 × 3

mm$^2$ crystal pixel to two segmented SiPM pixels.

This study presents findings, especially in terms of timing performance, obtained from the OctaSiPM coupled with BGO crystals, emphasizing the new possibilities offered by leveraging multiple timestamps from each detected annihilation event and the advantages of segmentation.

Segmented readout scheme

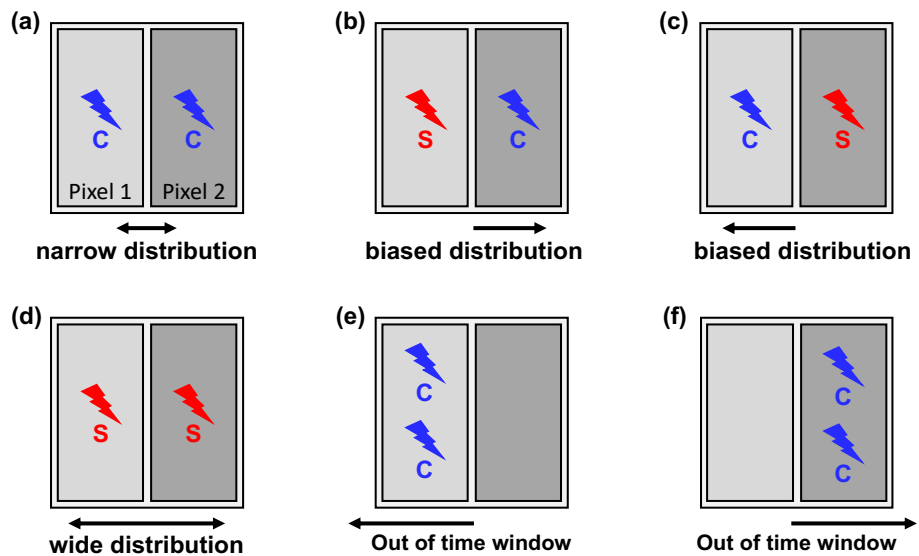

Figure 1. Detector triggering scenarios in segmented readout scheme and trigger time difference distribution. (a) Cherenkov – Cherenkov. (b) Scintillation – Cherenkov. (c) Scintillation – Scintillation. (d) Cherenkov – Scintillation. (e) No photon in a time window – Multiple photons. (f) Multiple photons – No photon in a time window.

In the proposed segmented readout scheme, the concept of 'trigger time difference' is introduced to characterize detected events. This is defined as the difference between the two timestamps obtained from the two SiPM pixels within a short time window at the onset of photon detection. The type of event can vary depending on whether each SiPM pixel is triggered by a Cherenkov photon or a scintillation photon. If both pixels are triggered by prompt Cherenkov photons (**Figure 1(a)**), the trigger time difference is expected to exhibit a narrow distribution centered around zero. In cases where one pixel is triggered by a scintillation photon and the other by a Cherenkov photon (**Figures 1(b) and 1(c)**), a biased distribution is anticipated, as the emission of scintillation photons is slower and less dense than that of Cherenkov photons. When both pixels are triggered by slow scintillation photons (**Figure 1(d)**), a wide distribution centered around zero is expected.

Assuming that events with small trigger time differences correspond to scenarios where a considerable number of prompt Cherenkov photons reached the SiPMs, a rough estimation of the initial photon density can be made based on trigger time differences. Here, the term 'initial photon density' refers to the number of photons detected within a very short, specified time interval immediately following the interaction of the annihilation photon.

However, for a more rigorous approach, if we consider a scenario where one of the two segmented pixels does not receive any photons within the time window after the other is triggered, as illustrated in **Figures 1(e) and 1(f)**, estimating the initial photon density solely through trigger time differences may pose challenges. Therefore, we will also introduce the concept of a trigger time window and the number of initially detected photons to characterize the annihilation event by integrating the very early part of the detector timing signal.

OctaSiPM and adaptive timestamp

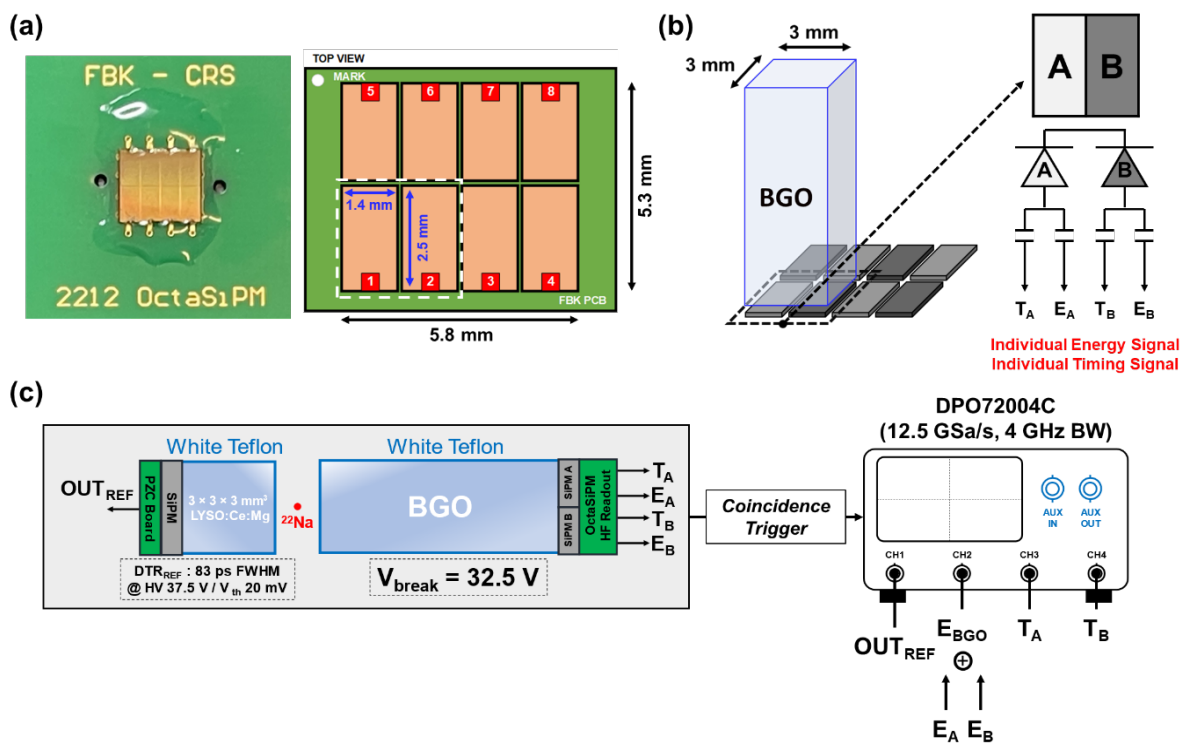

**Figure 2. Experimental instrumentation setup. (a) Photograph of an OctaSiPM (left) and dimension details (right). (b) A single BGO crystal was optically coupled to two pixels of an OctaSiPM. (c) Measurement setup for coincidence timing resolution detection of back-to-back annihilation photons from a $^{22}$Na point source.**

**Figure 2(a)** shows a picture of an OctaSiPM fabricated by FBK. The OctaSiPM was manufactured based on FBK's well-established near-ultraviolet high-density metal-filled-trench (NUV-HD-MT) technology **[18]**. At the level of an individual single photon avalanche diode (SPAD), the field configuration was optimized by using a metal mask to cover the edges, improving single SPAD signal characteristics for Cherenkov photon detection **[19]**. The SPADs were arranged with a pitch of 40 μm, and each SiPM pixel possesses an active area of 2.5 × 1.4 mm². The OctaSiPM is a 2 × 4 SiPM array, collectively offering a nominal active area of 5.3 × 5.9 mm².

Polished BGO pixels (Epic-crystal, Kunshan, China) with a cross-sectional area of 3 × 3 mm$^2$ and lengths of 5, 10, 15, and 20 mm were prepared. Each BGO pixel was optically coupled using silicon optical grease (BC-630, Saint-Gobain, France) to an OctaSiPM, covering two of its adjacent SiPM pixels (**Figure 2(b)**). For clarity and symmetry, we designated one as 'pixel A' and the other as 'pixel B' (*AB-segmented* configuration). The anode signal from each channel was fed into high-frequency electronics consisting of two broadband amplifiers (*HMC311SC70-8GHz,* Analog Devices, USA) for the timing channel and an operational amplifier (*AD8000,* Analog Devices, USA) for the energy channel **[24, 26, 32]**. This setup allowed us to obtain two timing signals and two energy signals from a BGO crystal pixel per 511 keV annihilation photon interaction. The trace lengths for the two timing signals were carefully routed to be equal while designing the detector readout board. The two energy signals were merged into a single energy signal and integrated to construct an energy histogram. The timestamps of each timing signal were derived through a leading-edge discrimination method, providing two timestamps ($T_A$, $T_B$) per 511 keV annihilation photon interaction. Here we define the trigger time difference as $T_A - T_B$. To select the earliest detected photon among those detected from the crystal, we employed an adaptive timestamp pickoff method, in which we select the earlier detected timestamp between $T_A$ and $T_B$ with a certain margin of $k$ on an event-by-event basis **(Eq. 1)**.

Adaptive timestamp method: $$T_{\text{Early}} = \begin{cases} T_A, & T_A - T_B < k \\ T_B, & T_A - T_B \geq k \end{cases}$$ (Eq. 1)

To assess the effect of segmentation on timing performance compared to a non-segmented SiPM, we prepared two comparison detector configurations. One configuration is a single SiPM pixel with an active area of 3 × 3 mm$^2$, matching a crystal pixel used in this study (*non-segmented* configuration). The single SiPM was fabricated using the same technologies as the OctaSiPM, except for the segmentation. The timestamp derived from this configuration will be referred to as '$T_{\text{nonseg}}$,' where 'nonseg' denotes the *non-segmented* SiPM. The other configuration was constructed by manually connecting the two anodes of the OctaSiPM pixels A and B, making it analogous to one single 3 × 3 mm² SiPM (*AB-connected* configuration). The timestamp obtained from this detector configuration will be denoted as '$T_{AB}$,' with 'AB' signifying the connection of SiPM pixels A and B. We conducted a comparative analysis between these two detector configurations and the one taking advantage of the segmented form of the OctaSiPM in terms of single-cell firing timing signal and timing performance.

Integration of the initial part of the timing signal

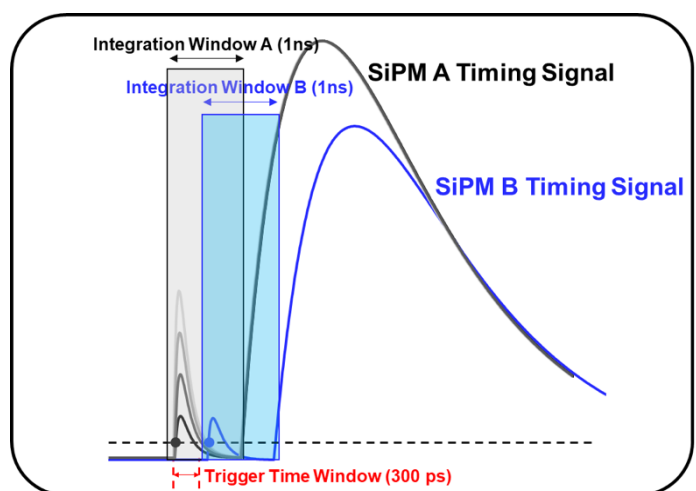

**Figure 3.** Integration window (1 ns) for the initial part of each timing signal and trigger time window. The trigger time window value was experimentally set to 300 ps, based on the point where the FWHM measured with the adaptive timestamp rapidly changed.

Given that we are working with BGO crystals functioning as a hybrid scintillator and Cherenkov radiator, it is worth noting that the temporal density of photons detected in the early part of the signal (initially detected photons) can impact affect the measured timing resolution. As Cherenkov photons are produced earlier than scintillation photons, the promptly detected photons are highly likely Cherenkov photons and contribute to improved timing performance. To characterize the initial photons, we applied an integration window of 1 ns to capture the very beginning part of each timing signal (**Figure 3**). We also introduced an additional time window, referred to as the trigger time window. If the signal detection in one pixel begins more than this trigger time window later than in another pixel, the initial photon number for the delayed pixel is considered zero, assuming that prompt photons, which contribute to improved timing performance, were not detected in the delayed pixel. The trigger time window value was experimentally set to 300 ps, based on the point where the FWHM measured with the adaptive timestamp rapidly changed, as shown in **Figure 5(c)**. This point is where events triggered by Cherenkov photons become dominant. For example, if a pixel B timestamp was 600 ps later than pixel A timestamp, then regardless of how many initial photons were detected from pixel B, we considered it as zero, indicating no prompt photons in pixel B.

Effect of segmentation on timing performance

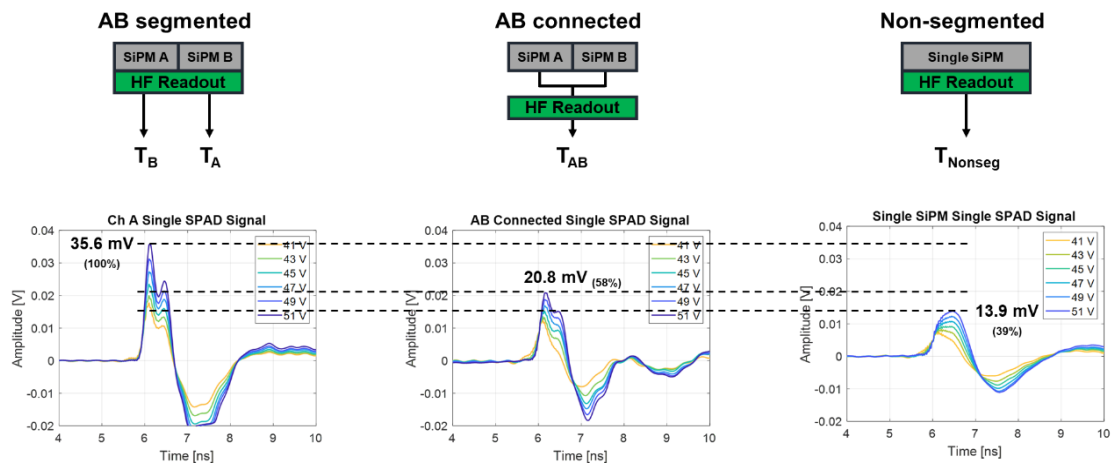

**Figure 4. Tested detector configurations and SPAD timing signals from each configuration with SiPM bias voltage ranging from 41V to 51V.**

The average signal from a fired SPAD cell at various SiPM bias voltages for each of the three detector configurations (*AB-segmented*, *AB-connected*, and *non-segmented*) is illustrated in **Figure 4**. When manually forming the connection between the two channels of an OctaSiPM (*AB-connected* configuration), the amplitude of the single-cell fired signal decreased by two-thirds compared to using the segmented pixel (*AB-segmented* configuration). When employing a non-segmented SiPM with a similar total active area (*Non-segmented* configuration), the decrease in signal amplitude was even more pronounced, reducing it to less than half of what was observed with the segmented configuration. Along with the amplitude, the slope of the rising edge of the signal also decreased. This result can be attributed to the heightened detector capacitance due to the increased parallel connection of SPADs, and this is expected to directly influence the timing performance. The difference in the shape of the signal and the further decreased amplitude in the *non-segmented* configuration compared to the *AB-connected* configuration can be ascribed to the difference in the connectors used between the SiPM and the readout board. Specifically, the OctaSiPM utilized a high-speed board-mounted header connector (ERM6 and ERF6 series, Samtec, USA), while the *non-segmented* SiPM employed regular pin connectors.

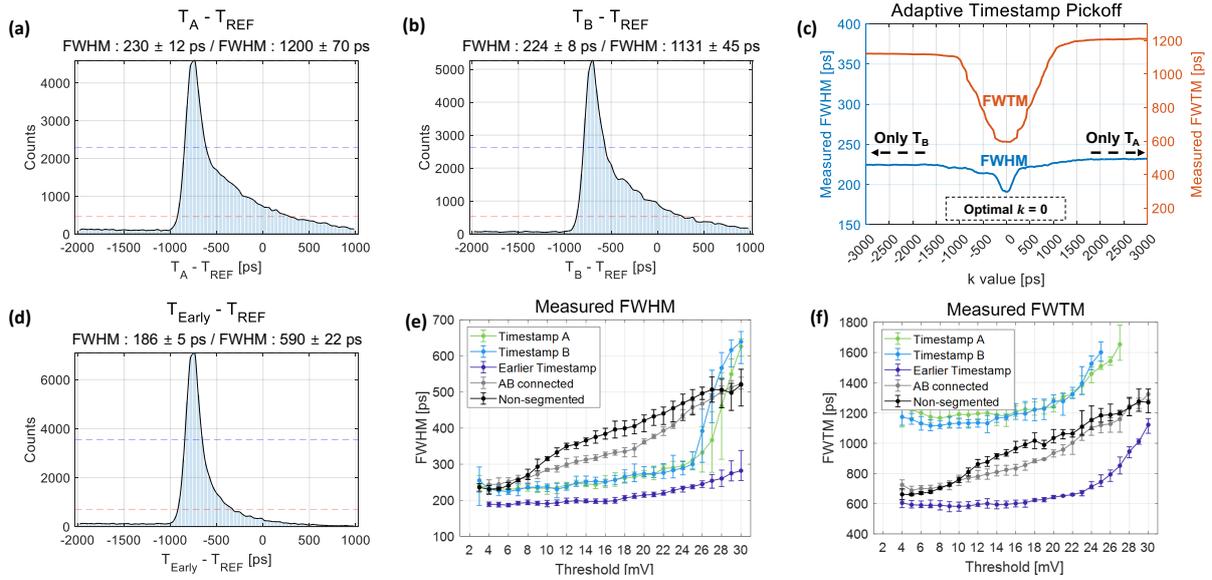

**Figure 5.** Measured timing histograms according to timestamp methods and the three different detector configurations. All results were at a SiPM bias voltage of 49 V. (a) Timestamp A. (b) Timestamp B. (c) Measured timing resolution values from adaptive timestamp method (d) Earlier timestamp ($k = 0$ in the adaptive timestamp method). (e) Measured FWHM values from three different detector configurations. (f) Measured FWTM values from three different detector configurations.

**Figures 5(a)** and **5(b)** show the time delay histogram between the reference detector and the 15 mm BGO detector coupled to an OctaSiPM when using a timestamp from either pixel A or B. Because each SiPM pixel covered only half of the crystal area, there is a significant loss of both scintillation photons and Cherenkov photons, leading to the elongation of the tails in the timing histograms: 230 ± 12 ps FWHM and 1200 ± 70 ps full-width-tenth-maximum (FWTM) for pixel A, and 224 ± 8 ps FWHM and 1131 ± 45 ps FWTM for pixel B. However, through the application of an adaptive timestamp pickoff method utilizing the same dataset, a notable enhancement in timing resolution was observed. The coincidence timing resolution (CTR) was measured and plotted as a function of $k$ (**Figure 5(c)**). The optimal $k$ value was determined to be 0, which is reasonable given the symmetry of the detector configuration. By setting the $k$ as 0, we observed significant improvements not only in an FWHM value of 186 ± 5 ps but also in an FWTM value of 590 ± 22 ps (**Figure 5(d)**).

**Figures 5(e)** and **5(f)** show a comparison of the measured timing resolution from the three detector configurations. First of all, the timestamps obtained only from pixels A or B in the *AB-segmented* configuration showed similar timing performance patterns (green and blue lines). By selecting the earlier detected timestamp between both timestamps on an event-by-event basis (Earlier timestamp; adaptive timestamp pickoff with the $k$ of 0), we could obtain a significant improvement in FWTM values as well as FWHM values (purple line).

In the case of the *AB-segmented* configuration, when using only one timestamp compared to selecting the earlier timestamp between pixels A and B, a notably extended tail and a higher FWTM value were observed. Timing resolution is proportional to $1/\sqrt{N}$, where N represents the total number of photons [33]. This prolonged tail occurs when events triggered by prompt photons are not properly detected because each SiPM pixel only covers half of the crystal area and N is halved, leading to a worsening of the timing resolution. Following a similar rationale, using the *AB-connected* configuration or *non-segmented* configuration was expected to yield better FWHM values compared to using only one side in the *AB-segmented* configuration. Nonetheless, similar optimum values were observed. This discrepancy was attributed to the segmentation effect, wherein the slope and amplitude of the SPAD signal increase, leading to a compensatory effect for the decrease in the number of received prompt photons. An abrupt and rapid degradation of the FWHM values was observed when employing only timestamps from either pixel A or B with a leading-edge threshold of around 26 mV, which is a typical SPAD signal amplitude of a segmented SiPM pixel, as shown in Figure 3. This is also explainable by the reduction in the detected prompt photons.

Timing performance changes with different ranges of trigger time differences

The distribution of trigger time differences between the two SiPM pixels A and B ($T_A - T_B$) is shown in **Figure 6(a)**. The distribution was spread widely between -3000 ps and 3000 ps, centered at 0 ps, due to the slow and sparse generation of scintillation photons. The discrete side peaks will be discussed in **Section 4.1**. Here, we will focus on the events whose trigger time difference falls within a certain range from $-\Delta T_k$ to $\Delta T_k$, where $\Delta T_k$ is defined as the threshold trigger time difference for events analyzed. **Figure 6(b)** shows the calculated CTR as a function of $\Delta T_k$, ranging from 50 ps to 3000 ps. As $\Delta T_k$ falls below a certain threshold, the timing resolution improved sharply and the ratio of FWTM to FWHM also approached its value for a single Gaussian distribution (1.83). This interesting observation implies that by reducing $\Delta T_k$, i.e., selecting events with small trigger time differences, we could effectively isolate events triggered by Cherenkov photons. The CTR results with different lengths of BGO crystal are summarized in **Supplementary Table 1**.

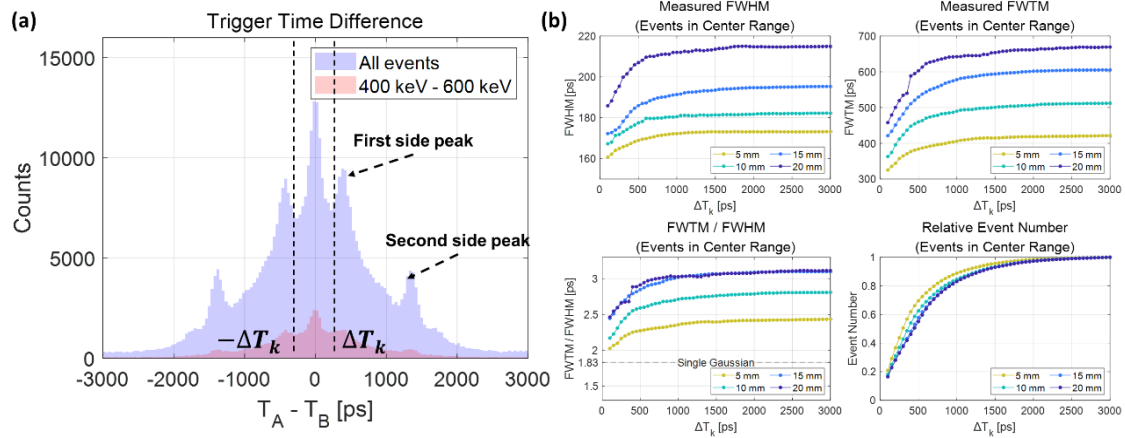

**Figure 6.** Event selection based on the trigger time difference ($T_A$-$T_B$). (a) Trigger time difference distribution. (b) Calculated coincidence timing resolution with different ranges of trigger time differences.

Integration of the early part of the timing signal

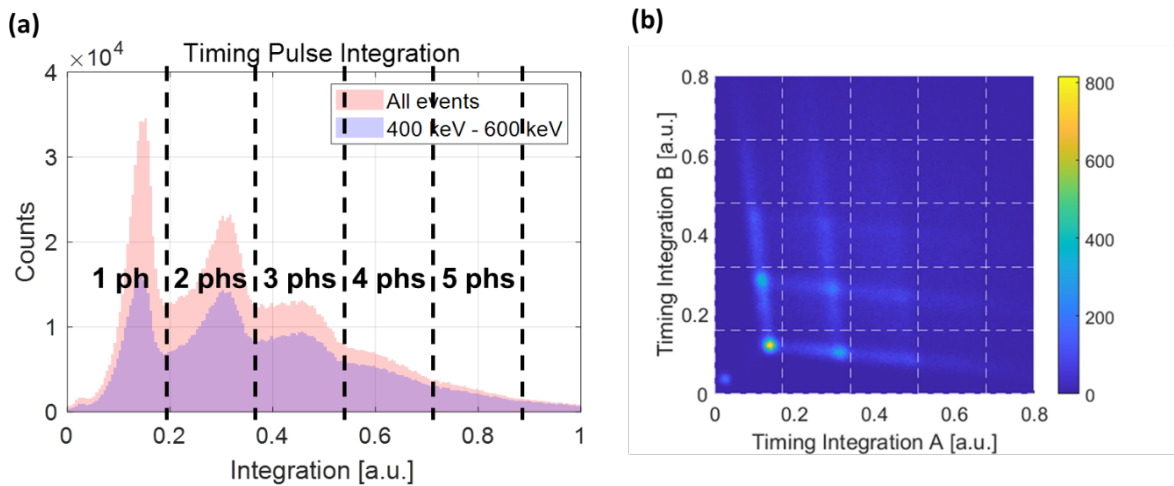

**Figure 7.** Counting initially detected photons. (a) Histogram of the integration of the early part of each timing signal from a SiPM pixel, along with the quantization of the timing signal integration value. (b) Two-dimensional density plot using integration values calculated from SiPM pixels A and B.

When we integrate the early part of each timing signal using a 1 ns integration window, as shown in **Figure 3**, the number of photons initially detected at each SiPM pixel (initially detected photons) can be measured, as shown in **Figure 7(a)**. By plotting a density plot with these integration values from SiPM pixels A and B, we obtained a grid-like distribution, as shown in **Figure 7(b)**. From this grid distribution, we could quantify the events corresponding to each grid by determining how many photons were initially detected by each SiPM pixel A and B.

The lowest horizontal grid line corresponds to events where multiple initial photons were detected in detector pixel A, while detector pixel B detected a single photon. Similarly, but conversely, the leftmost vertical grid line corresponds to events in which multiple initial photons were detected in detector pixel B, while detector pixel A detected a single photon. The lines between grid points result from the delayed arrival of initial photons due to differences in the photon transit length caused by internal reflections or the real production time difference among photons within the narrow integration window. Given that we have applied a narrow integration window for the early portion of a timing signal in each segmented SiPM pixel, the delayed arrival of photons results in the tail end of the detector response extending beyond the integration window, thereby reducing the signal integration value within that window (**Supplementary Figure 1**). Following the event quantification using the integration value, an additional filtering step with a trigger time difference window of 300 ps was applied, and the number of initially detected photons (or simply the initial photon number) for each SiPM pixel was finally determined.

Timing performance according to the initial photon number

**Figure 8(a)** represents the 2-D histogram of detected events based on the number of initial photons in SiPM pixels A and B. Aggregating this histogram diagonally represents the event count corresponding to the total number of initial photons in pixels A and B combined, resulting in the distribution shown in **Figure 8(b)**. The number of initial photons showed a mean value of 3.9 (represented by the blue bars). When exclusively considering events within the central range of trigger time differences ($\Delta T_k$ value of 300 ps), the distribution followed a Poisson distribution with a mean value of 5.3 (represented by the red bars and curve).

**Figure 8(c)** illustrates the distribution of the total initial photon numbers in pixels A and B for various lengths of BGO crystal pixels. This distribution followed a Poisson distribution, indicating a trend of increased average detection numbers due to reduced photon loss as the crystal length decreases.

**Figure 8(d)** shows the mean number of initial photons as a function of $\Delta T_k$. It was observed that events located in the region with low values of $\Delta T_k$ tend to exhibit a higher detection count. Once again, the shorter crystal always exhibited a higher count of initial photons due to reduced photon loss during light transport. The likelihood of an event having a smaller trigger time difference between the two pixels corresponds to a higher probability of it being associated with an event characterized by a higher initial photon density. This is indicative of the higher probability of a bigger bunch of Cherenkov photons having arrived. This observation could be considered in relation to the results obtained from **Figure 6(b)**. In essence, the improvement in timing performance and the convergence of the FWTM/FWHM ratio to that of a single Gaussian distribution with a narrower $\Delta T_k$ range could be explained by the preferential selection of events with higher initial photon density and an increased prevalence of purely Cherenkov-triggered events.

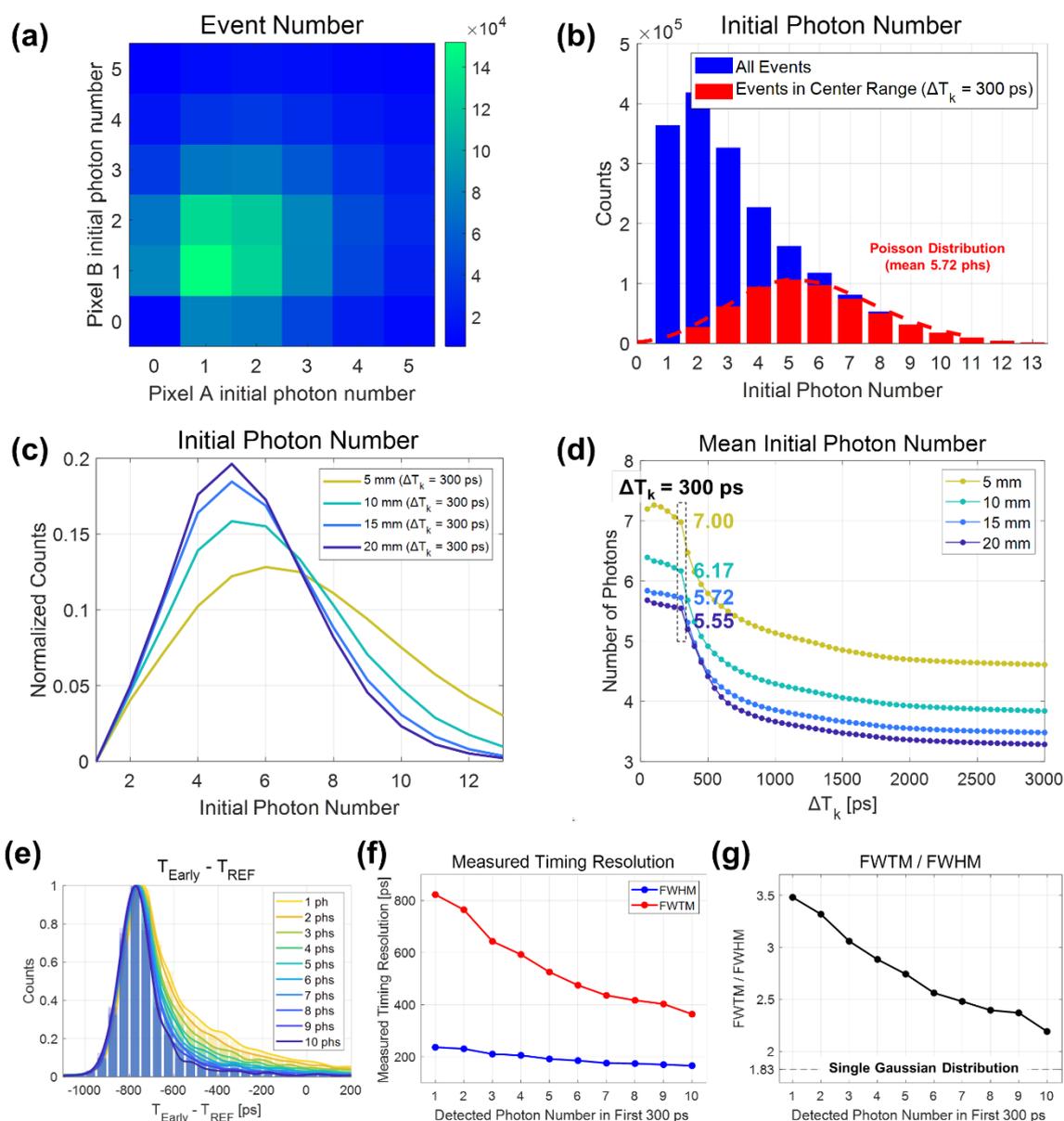

Figure 8. (a) Two-dimensional histogram of initial photon numbers in SiPM pixels A and B. (b) Distribution of the initial photon number for different trigger time difference ranges. (c) Distribution of the initial photon number for different BGO crystal lengths. (d) Distribution of the initial photon number as a function of $\Delta T_k$ range. (e) Coincidence time difference histograms according to the initial photon number. (f) Measured FWHM and FWTM values. (g) FWTM / FWHM ratio values.

We also calculated CTR values according to the number of initial photons. **Figures 8(e), 8(f), and 8(g)** illustrate the results obtained using a 15-mm long BGO crystal. An increase in the initial photon number is associated with a noticeable improvement in the measured timing resolution. To the best of our knowledge, this is the first direct experimental demonstration of the relationship between the initial photon count and timing resolution in BGO.

Analysis based on energy windows

It is known that a decreased energy deposition of an annihilation photon results in lower production of Cherenkov photons. **Supplementary Figure 2(a)** illustrates the normalized histogram of trigger time differences with varying lower energy cuts while maintaining a fixed upper energy cut at 600 keV. Indeed, as events with higher energy were selected, the number of events with large trigger time differences decreased, indicating an increased proportion of events with higher initial photon density. By setting the BGO energy window to 400 – 600 keV and comparing the trigger time difference histograms between events within this energy window and those outside of it, this characteristic becomes more pronounced (**Supplementary Figure 2(b)**). When calculating the mean number of initial photons for different BGO energy intervals (50 – 100 keV, 100 – 150 keV, …, 550 – 600 keV), the initial photon density indeed increased in higher energy events (**Supplementary Figure 2(c)**).

Histogram of trigger time differences between pixels A and B

The histogram of trigger time differences between pixels A and B exhibits symmetric and distinctly discrete side peaks on either side of the center. To identify the underlying causes of this phenomenon, we systematically varied several experimental parameters and observed trends. By adjusting the SiPM bias voltage, we observed that the relative height of the first side peaks diminished with a lower bias voltage (**Figure 9(a)**). When varying the length of the BGO crystal, the first side peaks shifted farther away from the center as the crystal length increased, as shown in **Figure 9(b)**, while the second side peaks, shown in **Figure 9(c)** at around ±1300 ps from the center, remained unchanged. When the threshold voltage for leading-edge discrimination was slightly increased, the second side peaks disappeared, while there were no discernible changes for the first peaks (**Figure

**9(c)**). Based on these observations, we inferred that the occurrence of the second side peaks was attributed to detector electronic crosstalk, whereas the first side peaks were associated with external optical crosstalk.

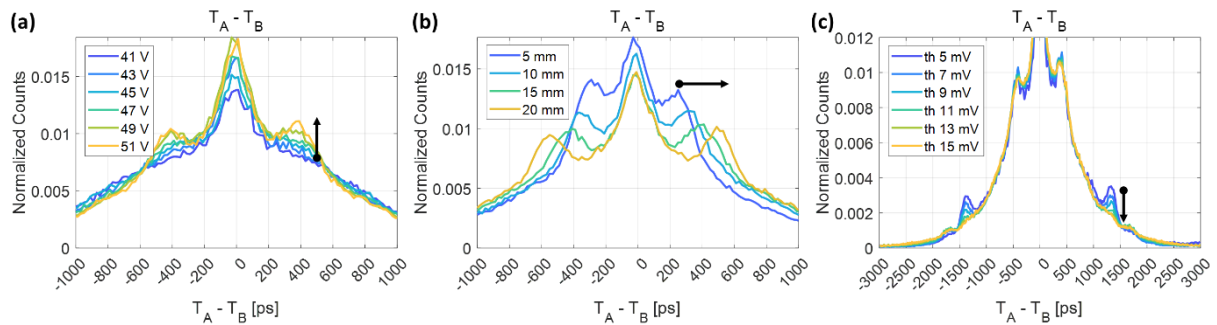

**Figure 9.** Observations on the trigger time difference distribution with different SiPM bias voltages (a), crystal lengths (b), and leading-edge threshold levels (c).

To assess the electronic crosstalk, we examined timing signal outputs from both segmented SiPM pixels A and B without any crystal coupling (**Supplementary Figure 3(a)**). This experiment revealed that the activation of one pixel results in the detection of a crosstalk signal with the opposite polarity within the other pixel (**Supplementary Figure 3(b)**). This electronic crosstalk was attributed to a current-sinking issue arising from the common cathode configuration of the OctaSiPM. Thus, if an adjacent pixel is not activated within a sufficiently short time after one pixel is fired, the observed electronic crosstalk can be captured when the trigger threshold level is set lower than the amplitude of the crosstalk signal. This phenomenon led to the presence of the second side peaks in the histogram of trigger time differences, as shown in **Figure 9(c)**. Because the amplitude of the crosstalk signal is smaller than that of a single SPAD-fired signal, the second side peaks vanish as the leading-edge discrimination threshold is increased.

To evaluate the impact of optical crosstalk within a crystal pixel, we performed an additional measurement using a BGO pixel with all surfaces, except for the OctaSiPM coupling surface, covered with black-painted PTFE tape. Compared to the white PTFE tape case, it was evident that the first side peaks in the trigger time difference histogram were significantly attenuated (**Supplementary Figure 4(a)**). As the events with a high density of initial photons decreased, the frequency of capturing the aforementioned electronic crosstalk increased, resulting in highly pronounced second side peaks **(Supplementary Figure 4(b))**. Indeed, when the crystal was wrapped with a black reflector, a significant portion of the emitted photons were absorbed by the black reflector, reducing the number of initial photons (blue bars in **Supplementary Figure 4(c)**). Moreover, the average number of initial photons in the central trigger time difference range (ΔT$_k$ value of 300 ps) also decreased from 5.72 with the white reflector to 3.45 with the black reflector (red bars in **Supplementary Figure 4(c)**). Similar to the results with the white reflector configuration shown in **Figure 8(f)**, the CTR improved along with the number of initial photons (**Supplementary Figure 4(d)**). The measured FWHM was slightly improved compared to the white reflector configuration (**Supplementary Figure 4(e)**) due to the reduced time dispersion of prompt photons by the black reflector **[13, 17]**. However, the absorption by the black tape also reduced the detection of Cherenkov photons at the same time, causing considerable degradation in the FWTM values (**Supplementary Figure 4(f)**).

Coupling an OctaSiPM to an LYSO:(Ce,Mg) crystal

We also performed a comparative test with a 3 × 3 × 20 mm$^3$ LYSO:(Ce,Mg) crystal pixel using the same setup. Similar to the BGO, within the OctaSiPM detector configuration, selecting the earlier detected timestamp between SiPM pixels A and B (152 ps) exhibited superior CTR compared to using timestamps from only one of them (194 ps and 180 ps, respectively) (**Supplementary Figure 5(a)**). Interestingly, as shown in the dotted boxes in **Supplementary Figure 5(a)**, when using timestamps from either pixel A or B alone, a minor discrepancy with the Gaussian fitting function was observed, attributable to photon loss. However, this issue was resolved when using the earlier timestamps. Comparing it with the non-segmented SiPM configuration (**Supplementary Figure 5(b)**), it was challenging to discern a significant improvement in timing resolution between the non-segmented SiPM (158 ps) and the segmented SiPM (152 ps) unlike the BGO. This difference from the BGO results is attributed to the fact that, in the case of LYSO:(Ce,Mg), the early temporal density of scintillation photons is notably high, and it possesses high photon statistics. Consequently, the detector segmentation effect was less pronounced for LYSO:(Ce,Mg) compared to the BGO, where the detection of the Cherenkov photon had a more pronounced impact.

Event classification for multi-kernel reconstruction approach

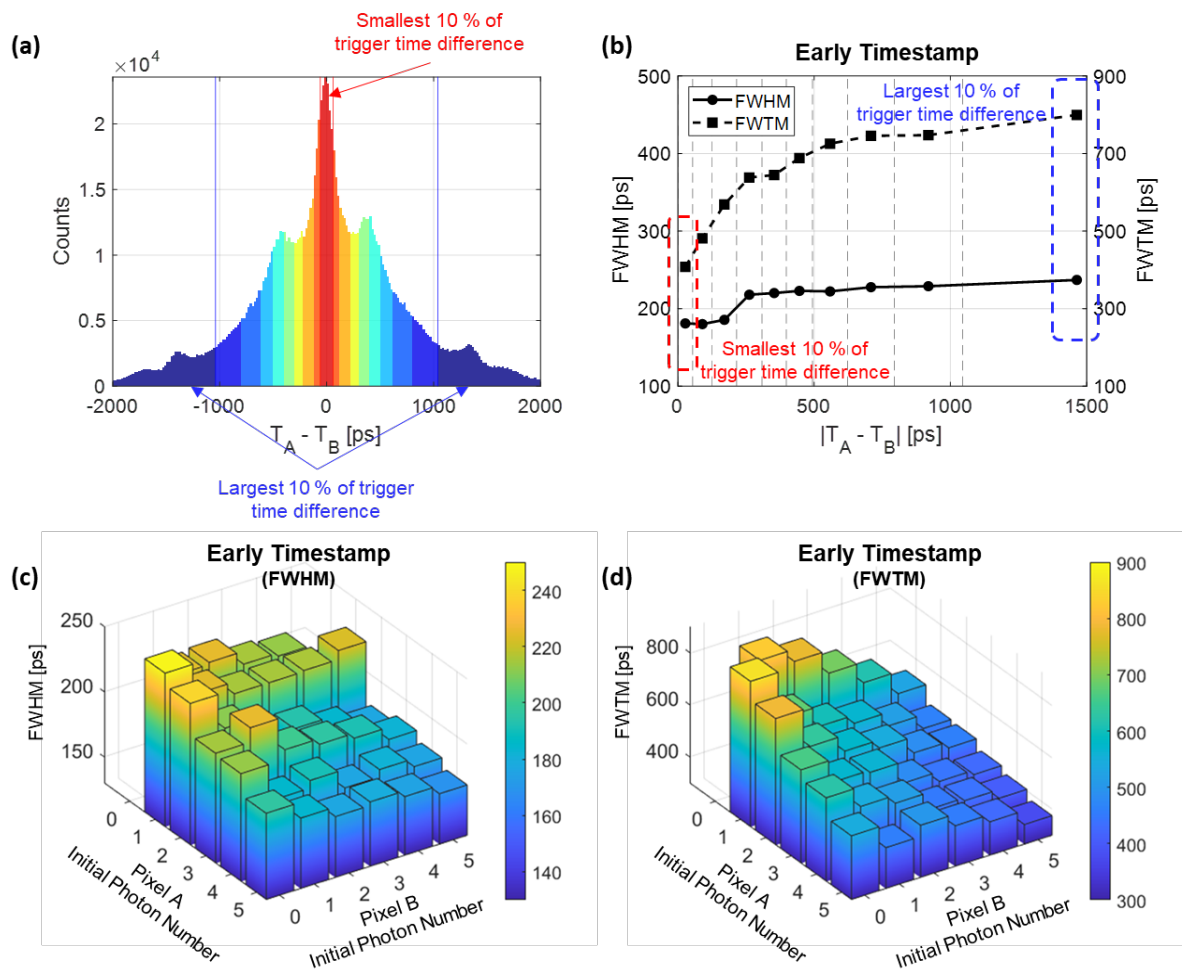

Figure 10. Event classifications using an OctaSiPM: one is based on the trigger time difference (a) and calculated timing resolution values of event classes (b), and the other is based on the number of initial photons in each segmented SiPM, resulting in various FWHM (c) and FWTM (d) values of event classes.

For BGO TOF PET detectors based on detecting Cherenkov photons, the variable detection of Cherenkov photons underlies event-to-event timing fluctuations. Therefore, the detection process does not offer a singular timing resolution; instead, it provides a range of timing resolutions. In this study, we confirmed that events with a smaller trigger time difference between the two segmented SiPM pixels exhibit better timing resolution because they tend to show a higher initial photon density. Furthermore, considering the finding that selecting events with smaller trigger time differences allows for higher probabilities of identifying events triggered purely by Cherenkov photons, one can contemplate employing trigger time differences for event classifications and adopting a multi-kernel reconstruction approach **[28]**. For example, we categorized all selected photopeak events into 10 groups based on their absolute trigger time differences ($|T_A - T_B|$) as shown in **Figure 10(a)**. Each color zone represents 10% of the events. Calculating the FWHM and FWTM for each group revealed distinct timing kernels suitable for each event group (**Figure 10(b)**). Particularly noteworthy is the FWHM, where a significant change occurs as events transition from the third group to the fourth group. In contrast, the FWTM values exhibit a pronounced initial variation, gradually leveling off in the latter groups. Also, incorporating the number of initial photons alongside trigger time differences could allow for another approach to classify events, as illustrated in **Figures 10(c)** and **10(d)**.

In timing measurements utilizing Cherenkov radiation under photon-poor conditions, the necessity for faster and larger SPAD signals becomes more pronounced. Additionally, to address the challenges stemming from the relatively low production of Cherenkov photons in BGO, strategies have been studied to overcome the different timing spread caused by photon fluctuations. In this study, a segmentation approach was employed to couple two smaller SiPMs to a single crystal. This not only improved the SPAD signal of SiPMs but also revealed a significant correlation between the time difference in detection between the two neighboring SiPM pixels and the measured timing performance. Taking advantage of the reduced detector capacitance and the diminished crystal coverage per SiPM pixel, the study demonstrated the feasibility of counting the number of detected photons in the early part of the timing signal. Ultimately, this approach allowed us to characterize each event by its initial photon density, which is significantly important in timing performance, simply by analyzing the time difference in detection between two adjacent SiPMs. This approach opens up new avenues for applications requiring precise timing and photon counting, extending beyond medical imaging to encompass high-energy and optical physics.

**Methods**

**Figure 2(c)** illustrates an experimental arrangement for coincidence event measurement. The coincidence detection of back-to-back annihilation photons from a $^{22}$Na point source was carried out between a reference detector and a BGO detector. The reference detector consisted of a polished 3 × 3 × 3 mm$^3$ LYSO:(Ce,Mg) (Taiwan Applied Crystals, Taiwan) pixel, wrapped with white polytetrafluoroethylene (PTFE) tape. The reference crystal pixel was optically coupled to a 3 × 3 mm$^2$ NUV-HD-MT SiPM (FBK, Trento, Italy) with silicone optical grease (BC-630) and read out through a pole-zero-cancelation (PZC) circuit **[34]**. The detector timing resolution of the reference detector was measured and derived as 83 ps full-width at half-maximum (FWHM) using three different detector pairs and a system of linear equations **[35]**.

The output signals were digitized using a four-channel oscilloscope (DPO72004C, Tektronix, USA) with a sampling rate of 12.5 GSPS and a bandwidth of 4 GHz. One of the oscilloscope channels was dedicated to digitizing output signals from the reference detector, while the remaining three channels were allocated to record output signals from the BGO detector. This configuration allowed us to allocate one consolidated energy signal and two distinct timing signals to the three channels, respectively. Events falling within the energy window of 400-600 keV for both the reference and BGO detectors were used in the timing performance analysis.

**Data and code availability**

The data that support the findings of this study are available from the corresponding author upon reasonable request.

**Acknowledgments**

This work was supported by a National Institute of Health grant R01 EB029633. The work of Minseok Yi was supported in part by the grant of the Korea Health Technology R&D Project through the Korea Health Industry Development Institute (KHIDI), funded by the Ministry of Health & Welfare, Republic of Korea (grant number: HI19C1352).

**Author information**

Authors and Affiliations

**Department of Biomedical Engineering, University of California, Davis, USA**

Minseok Yi, Daehee Lee, Simon R. Cherry & Sun Il Kwon

**Interdisciplinary Program in Bioengineering, College of Engineering, Seoul National Graduate School, Republic of Korea**

Minseok Yi & Jae Sung Lee

**Integrated Major in Innovative Medical Science, Seoul National Graduate School, Republic of Korea**

Minseok Yi & Jae Sung Lee

**Fondazione Bruno Kessler, Italy**

Alberto Gola, Stefano Merzi & Michele Penna

**Brightonix Imaging Inc., Republic of Korea**

Jae Sung Lee

Corresponding author

Correspondence to SUN IL KWON (sunkwon@ucdavis.edu)

**Supplementary Table 1. Coincidence timing resolution values measured using different BGO crystal lengths.**

| Length | 5 mm | | 10 mm | | 15 mm | | 20 mm | |
|---|---|---|---|---|---|---|---|---|
| [ps] | FWHM | FWTM | FWHM | FWTM | FWHM | FWTM | FWHM | FWTM |
| $T_A$ | 185 ± 3 | 826 ± 36 | 210 ± 11 | 1084 ± 56 | 230 ± 12 | 1200 ± 70 | 255 ± 14 | 1257 ± 53 |
| $T_B$ | 189 ± 4 | 861 ± 43 | 203 ± 13 | 1005 ± 22 | 224 ± 8 | 1131 ± 45 | 250 ± 13 | 1277 ± 44 |
| $T_{Early}$ All Events | 167 ± 3 | 423 ± 27 | 175 ± 5 | 502 ± 17 | 186 ± 5 | 590 ± 22 | 204 ± 5 | 648 ± 28 |
| $T_{Early}$ ($\Delta T_k = 300\ ps$) | 161 ± 6 | 359 ± 11 | 164 ± 4 | 399 ± 16 | 183 ± 8 | 500 ± 28 | 181 ± 10 | 500 ± 20 |

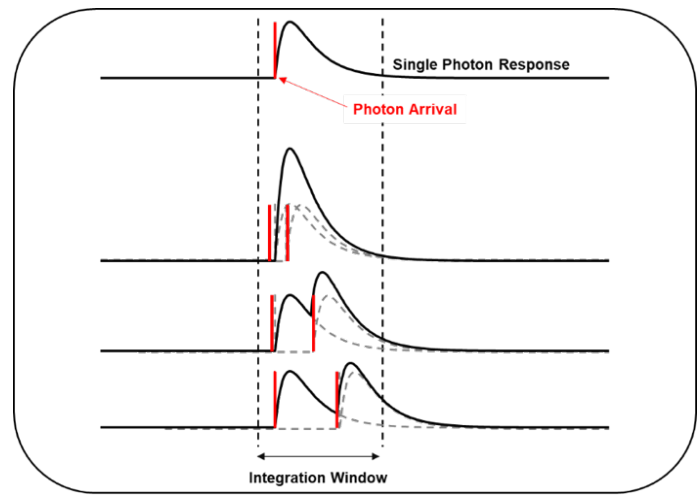

**Supplementary Figure 1. The effect of photon arrival delay on the integration value.**

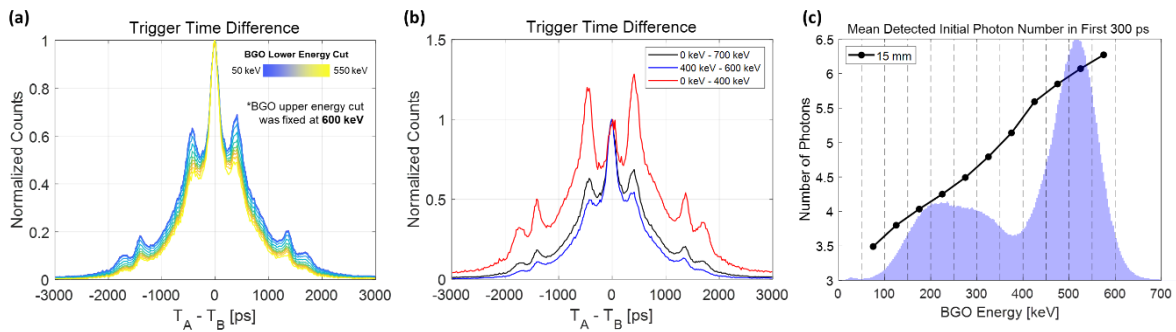

**Supplementary Figure 2. Distributions of trigger time differences with varying energy lower limits. (a) With a fixed 600 keV upper limit (b) Photopeak (400 – 700 keV) and scatter (0 – 400 keV) energy regions. (c) The number of initial photons for different energy intervals (50 – 100 keV, 100 – 150 keV, …, 550 – 600 keV).**

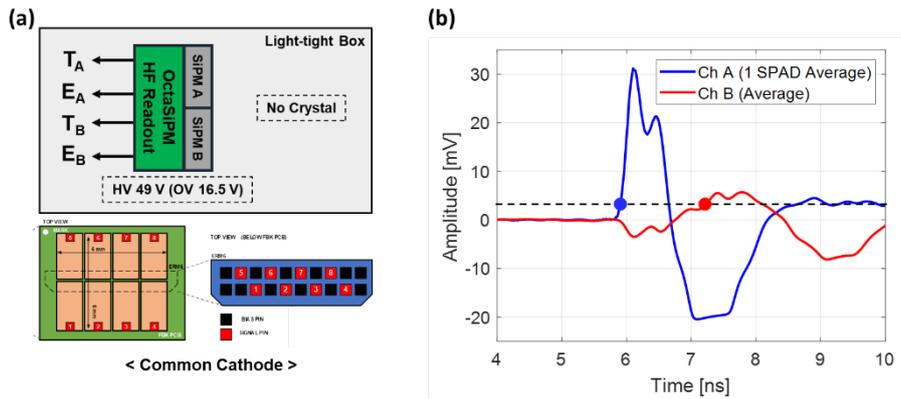

**Supplementary Figure 3.** Electronic crosstalk due to the common cathode configuration.

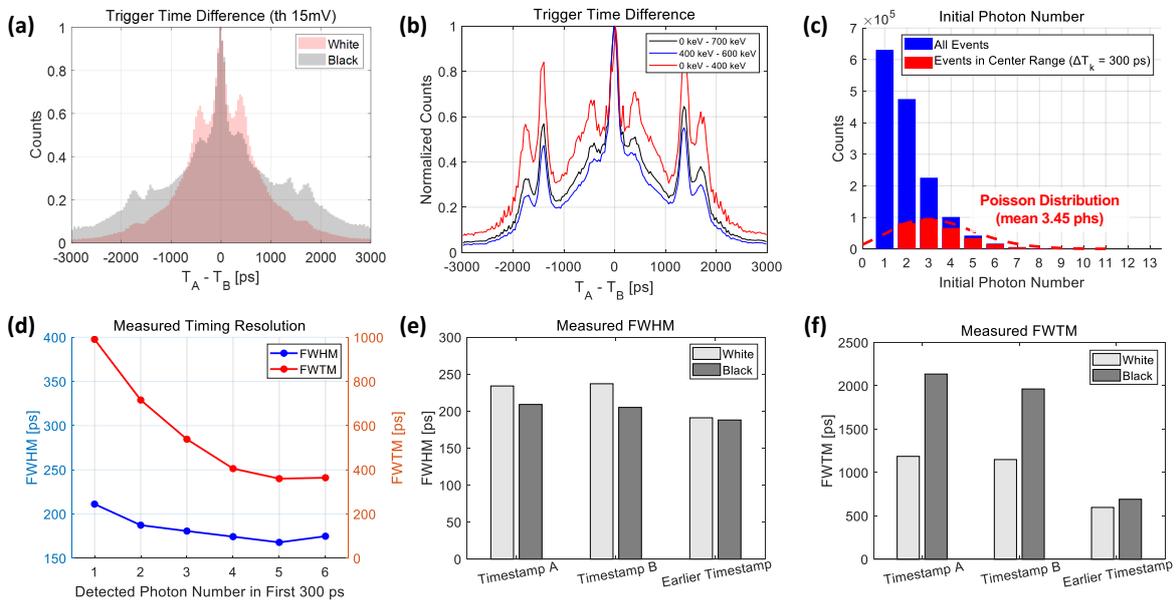

**Supplementary Figure 4.** Black reflector results.

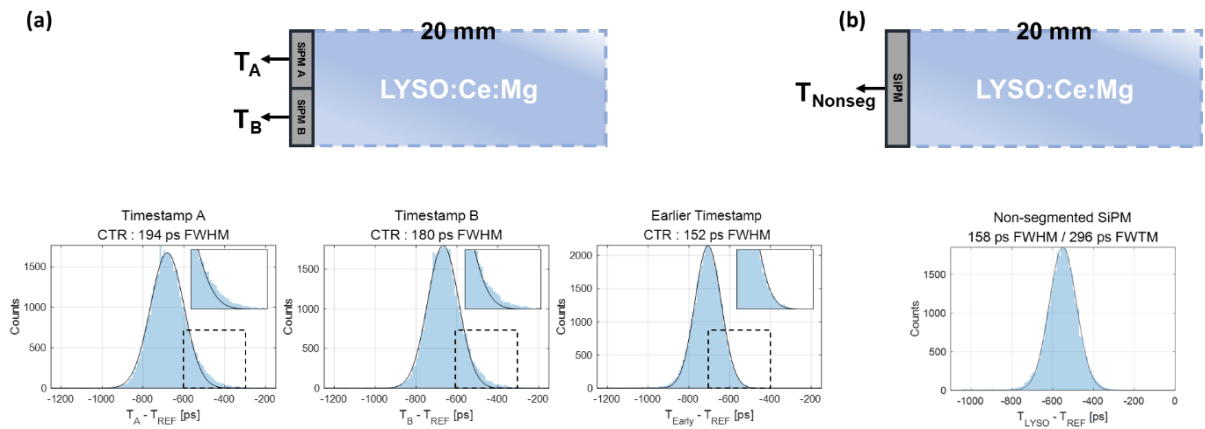

**Supplementary Figure 5.** Measurements from LYSO:(Ce,Mg) coupled to an OctaSiPM (a) and a non-segmented SiPM (b).